\documentclass[prl,aps,amsfonts,amsmath,amssymb,nofootinbib,twocolumn,superscriptaddress]{revtex4-2}
\usepackage{graphicx}
\usepackage{bm}
\usepackage{float}
\usepackage{xcolor}
\usepackage{hyperref}
\graphicspath{{figures/}}
\begin{document}
\title{Absence of two-phonon quasi-elastic scattering in the normal state of doped--SrTiO$_3$ by THz pump-probe spectroscopy}
\author{K. Santhosh Kumar}
\affiliation{William H. Miller III, Department of Physics and Astronomy, The Johns Hopkins University, Baltimore, Maryland 21218, USA}
\author{David Barbalas}
\affiliation{William H. Miller III, Department of Physics and Astronomy, The Johns Hopkins University, Baltimore, Maryland 21218, USA}
\author{Rishi Bhandia}
\affiliation{William H. Miller III, Department of Physics and Astronomy, The Johns Hopkins University, Baltimore, Maryland 21218, USA}
\author{Dooyong Lee}
\affiliation{Department of Chemical Engineering and Materials Science, University of Minnesota, Minneapolis, Minnesota 55455, USA.}
\affiliation{Department of Physics Education, Kyungpook National University, 80, Daehak-ro, Buk-gu, Daegu 41566, South Korea.}
\author{Shivasheesh Varshney}
\affiliation{Department of Chemical Engineering and Materials Science, University of Minnesota, Minneapolis, Minnesota 55455, USA.}
\author{Bharat Jalan}
\affiliation{Department of Chemical Engineering and Materials Science, University of Minnesota, Minneapolis, Minnesota 55455, USA.}
\author{N. P. Armitage}
\email{npa@jhu.edu}
\affiliation{William H. Miller III, Department of Physics and Astronomy, The Johns Hopkins University, Baltimore, Maryland 21218, USA}
\affiliation{Canadian Institute for Advanced Research, Toronto, Ontario M5G 1Z8, Canada}

\begin{abstract}

Multi-pulse nonlinear THz spectroscopies enable a new understanding of interacting metallic systems via their sensitivity to novel correlation functions. Here, we investigated the THz nonlinear properties of the dilute metallic phase of doped-SrTiO$_3$ thin films using nonlinear terahertz 2D coherent spectroscopy. We observed a large $\chi^{(3)}$ response in the low temperature region where the dc electrical resistivity follows a T$^2$-dependence.  This is largely a pump-probe response, which we find is governed by a single energy relaxation rate that is much smaller at all temperatures than the momentum relaxation rates obtained from the optical conductivity.  This indicates that the processes that dominate the resistive scattering are not the same as those that remove energy from the electronic system.   Moreover the fact that the energy relaxation rate is an increasing function of temperature indicates that the excitations that do carry away energy from the electronic system cannot be considered as quasi-elastic and as such soft two-phonon electron scattering does not play a major role in the physics as proposed.  This indicates that these materials' resistive T$^2$ scattering likely originates in electron-electron interactions despite the very small Fermi wave vectors at the lowest dopings.

\end{abstract}

\maketitle

The temperature dependence of a metal's electrical resistivity is governed by various types of charge scattering. The low-temperature resistivity of many metals exhibits a dependence $\rho(T) = \rho_0+A$T$^2$, where $A$ is a material specific parameter and $\rho_0$ is the residual resistivity.  T$^2$-resistivity arises in conventional metals due to momentum-relaxing electron-electron scattering of either the umklapp variety or Baber's mechanism of scattering between different anisotropic parts of a Fermi surface~\cite{Landau,Baber}.  Normal electron-electron scattering does not relax momentum.  Materials with quadratic temperature dependent resistivity have been observed frequently in many interacting fermionic systems, including Sr$_2$RuO$_4$~\cite{PhysRevLett.120.076602, PhysRevLett.113.087404}, heavy fermion compounds~\cite{PhysRevLett.57.1955}, iron pnictides~\cite{PhysRevLett.109.187005, PhysRevB.89.220509}, and even the thermal resistivity in $^3$He~\cite{behnia2022origin}.  The interaction itself can be of the direct Coulomb interaction or phonon mediated variety~\cite{macdonald1980electron,van2011common,klimin2012microscopic}.  T$^2$ scattering is quite generic as it arises from phase space considerations of fermion scattering near E$_F$.  However, there are also many metals that show anomalous non-Fermi liquid properties in their transport at low temperatures e.g. exponents of temperature's power law behavior less than two ~\cite{schofield1999non,stewart2001non}.  In this regard it is important to investigate the robustness of Fermi liquid states and departures from canonical behavior.

Doped-SrTiO${_3}$ may be said to be the first unconventional oxide superconductor and possesses a number of exceptional properties~\cite{PhysRevLett.12.474}. It exhibits superconductivity in the doping range of $\sim 3 \times$ 10$^{17}$/cm$^3$- 10$^{21}$/cm$^3$ making it by far the most dilute carrier density superconductor~\cite{PhysRevLett.12.474, GASTIASORO2020168107,collignon2019metallicity}. The standard BCS theory cannot simply explain the mechanism of superconductivity owing to the inversion of the relative scales of the Fermi and Debye energies~\cite{collignon2019metallicity}.  It has long been suspected that there is a connection between the superconductivity in doped materials and aborted ferroelectricity in the parent compound~\cite{collignon2019metallicity}.  Although undoped-SrTiO${_3}$ approaches a ferroelectric state at low temperature, zero-point quantum fluctuations ultimately prevent long range order.  Its optical spectra is dominated by a soft transverse optical (TO) phonon whose frequency does not completely go to zero at the lowest temperatures, but instead stalls out around 1.2 meV (0.29 THz)~\cite{yamada1969neutron,fleury1968electric,van2008electron}.  This low frequency phonon gives a static dielectric constant of greater than 20,000 at low temperature~\cite{weaver1959dielectric}, which undoubtedly must influence the effective interactions.

The normal state of doped-SrTiO$_3$ is also interesting.  It exhibits a T$^2$ dependent electrical resistivity, the coefficient of which evolves smoothly from a high density regime where multiple anisotropic bands are occupied to a low density regime where a single isotropic band is occupied~\cite{mikheev2016carrier,lin2015scalable,behnia2022origin}.  Although T$^2$ resistivity is familiar in interacting metals, it remains to be understood how a single isotropic Fermi surface with $k_F$ too small to engage in umklapp processes can give electron-electron momentum non-conserving scattering.  There have been proposals for a non-electronic origin of the T$^2$ scattering~\cite{kumar2021quasiparticle,yu2022theory}.  For electrons near the Brillouin zone center, scattering from a single soft TO phonon is suppressed in a single-band system with weak spin-orbit interaction.  However, two-phonon scattering is allowed and has been proposed long ago as a mechanism for superconductivity~\cite{ngai1974two}.  For temperatures above the soft-mode frequency its contribution to the resistivity is expected to scale with the occupation number of phonons {\it squared}, i.e., as T$^2$~\cite{kumar2021quasiparticle}.  Two-phonon scattering provides a natural scenario for non-electronic T$^2$ resistivity in a soft mode system.  However, it is not clear that doped SrTiO$_3$ (and even more so the doped {\it films} of the present study) have a TO phonon soft enough to be in the equipartition regime.  Moreover, a similar T$^2$ resistivity has also been seen in the low density metal Bi$_2$O$_2$Se, which has no soft phonons~\cite{wang2020t}.  Additionally, the observation of a Kadowaki-Woods-like scaling of the coefficient of the resistivity being $ \propto E_F^{-2}$ and a T$^2$-dependent thermal resistivity gives evidence for an electronic contribution to the scattering~\cite{lin2015scalable, PhysRevLett.131.016301}. However, an unconventional scaling of the relative sizes of the T$^2$ and $\omega^2$ dependent scattering rates in our recent THz conductivity measurements indicates physics beyond conventional Fermi liquid theory~\cite{kumar2024anomalous}. Therefore, many issues remain regarding the origin of T$^2$ resistivity in doped-SrTiO${_3}$.

In the present work, we implemented nonlinear THz 2D coherent spectroscopy (2DCS) to understand the {\it energy} relaxation rate in the dilute metallic phase of doped-SrTiO$_3$ thin films. We observed a $\chi^{(3)}$ response, which is largest in the low temperature regime where dc electrical resistivity follows a T$^2$-dependence (below 120 K).  The relaxation of the THz pump-THz probe response is governed by a single relaxation time, which we can relate to the rate that energy leaves the electronic system.  We find the energy relaxation rate is an increasing function of temperature, and in magnitude is much smaller than the momentum relaxation rate observed in the Drude optical conductivity.   Its relative scale shows that scatterings that remove momentum from the electronic system largely do not remove energy.  Moreover, the fact that the energy relaxation is an increasing function of temperature rules out a dominant contribution of scattering from two soft transverse optical phonons at temperatures above their characteristic energy.   At such temperatures scattering should be largely elastic~\cite{allen_theory_1987} and the expectation is that the energy relaxation rates would decrease as 1/T.  From this we can conclude that it is likely that the T$^2$ resistivity appears to have an electronic origin, which remains to be understood.

\begin{figure}[t]
    \centering
    \includegraphics[width = 9cm]{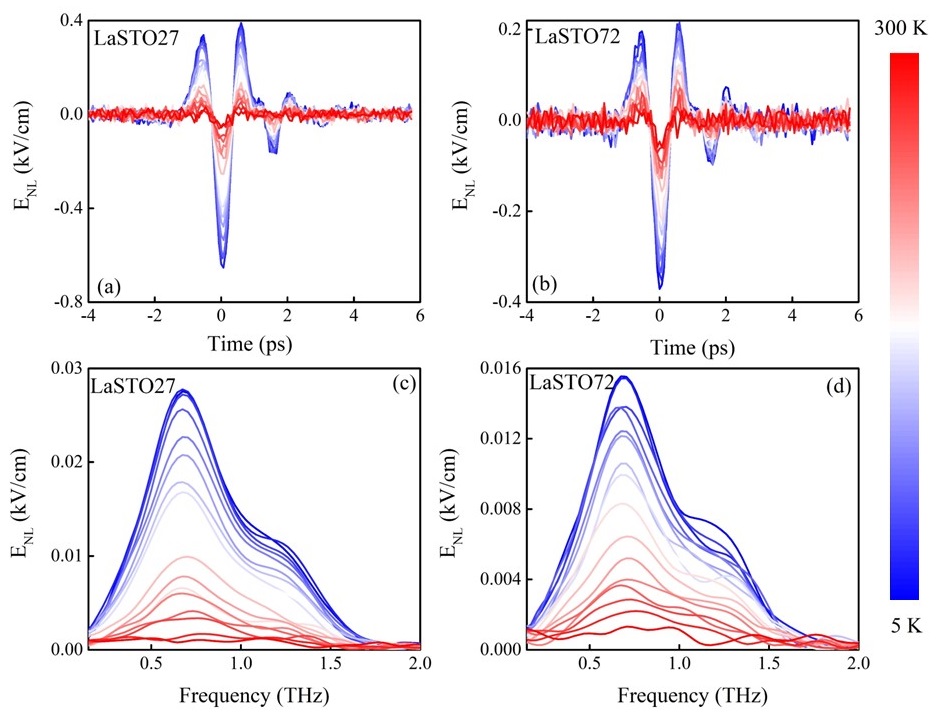}
    \caption{The THz nonlinear response of doped-SrTiO$_3$ thin films in the temperature range 5-300K at two different concentrations. The time traces of the nonlinear signal $E_{NL}(t, \tau =0)$ for (a) LaSTO27 ($2.7 \times 10^{20}/$cm$^3$) and (b) LaSTO72 ($7.2 \times 10^{20}/$cm$^3$), respectively. Fourier transform of the signal in panels (a) and (b) is shown in (c) LaSTO27 and (d) LaSTO72, respectively.}
    \label{Chi3}
\end{figure}

\begin{figure}[h]
    \centering
    \includegraphics[width = 9cm]{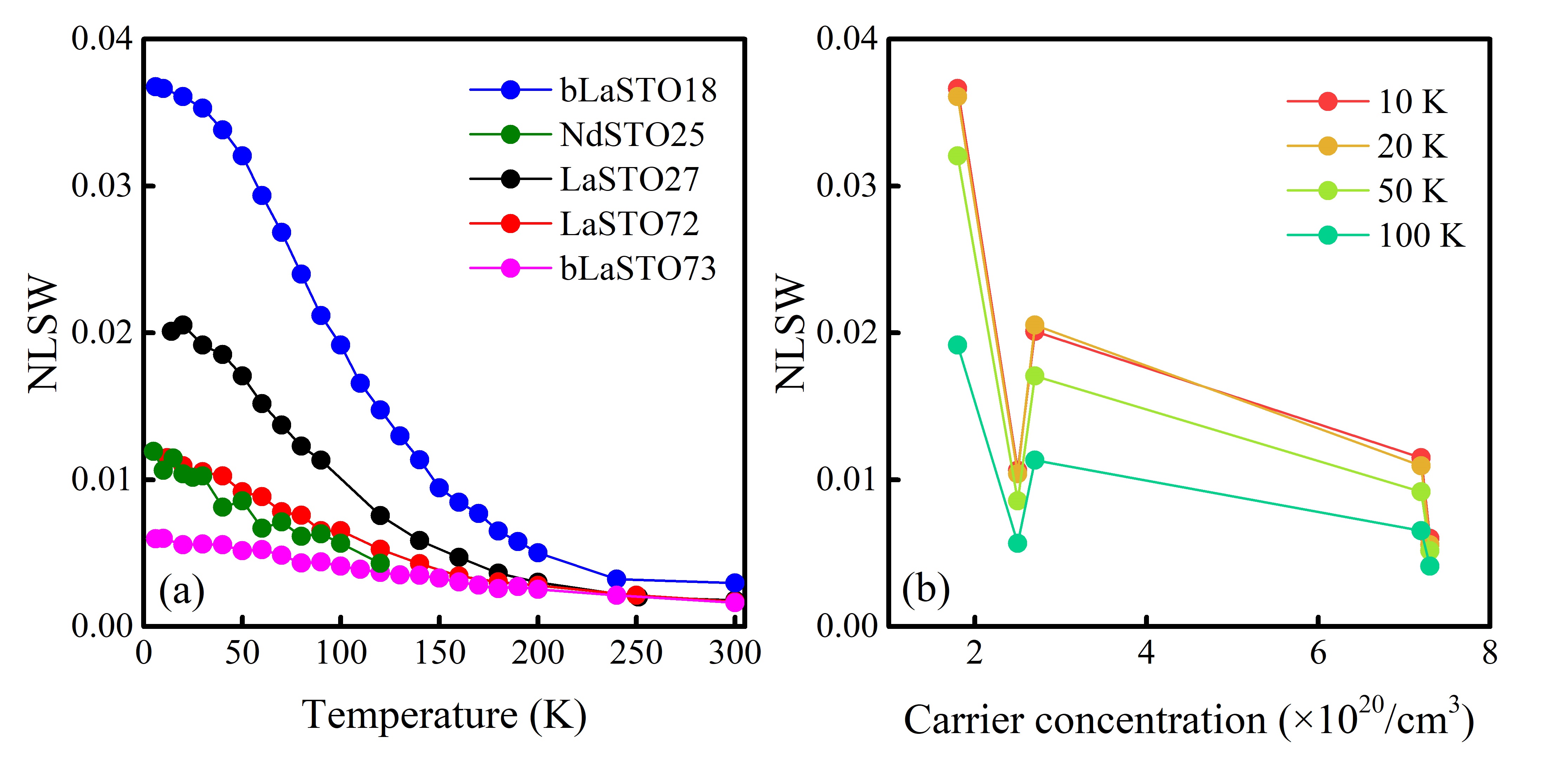}
    \caption{(a) Temperature dependent THz nonlinear spectral weight (NLSW) integrated over the range 0.2-1.4THz for doped-SrTiO$_3$ thin films at various carrier concentration bLaSTO18 (1.8 $\times$ 10$^{20}$/cm$^3$, 100 nm thickness), NdSTO25 (2.5 $\times$ 10$^{20}$/cm$^3$, 194 nm thickness), LaSTO27 (2.7 $\times$ 10$^{20}$/cm$^3$, 100 nm thickness), LaSTO72 (7.2 $\times$ 10$^{20}$/cm$^3$, 66 nm thickness), and bLaSTO73 (7.3 $\times$ 10$^{20}$/cm$^3$, 100 nm thickness). (b) THz nonlinear spectral weight from panel (a) is plotted as a function of carrier concentration at various temperatures. }
    \label{Chi3spec}
\end{figure}

\begin{figure*}[t]
    \centering
    \includegraphics[width = 16cm]{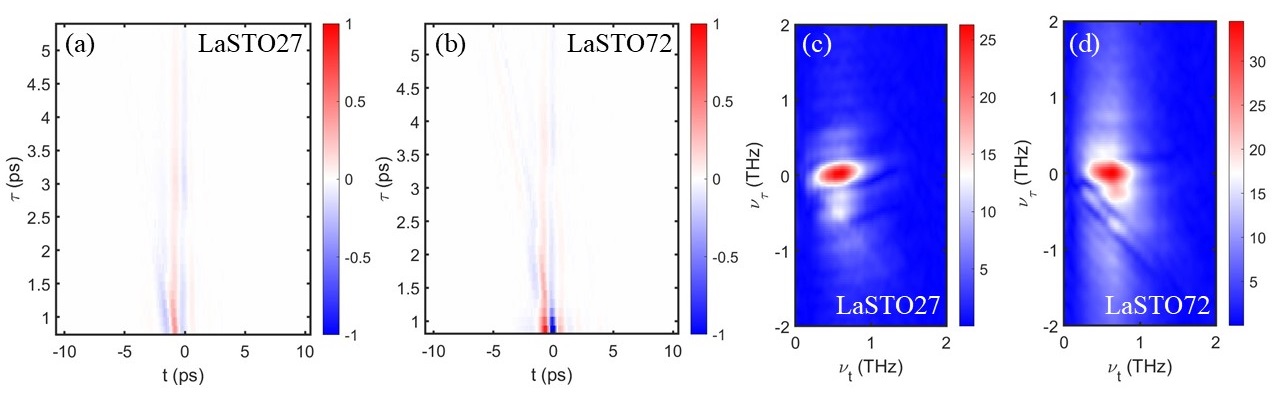}
    \caption{(a), (b) Time domain 2D THz response for time traces for $t$ and $\tau>0.7ps$ of LaSTO27 and LaSTO72  samples.  (c) and (d) Fourier transforms of time-domain traces. }
    \label{TwoD}
\end{figure*}

Epitaxial La-and Nd-doped SrTiO$_3$ thin films were grown on (001) oriented (LaAlO$_3$)$_{0.3}$(Sr$_2$AlTaO$_6$)$_{0.7}$ (LSAT) single-crystal substrates using hybrid molecular beam epitaxy. Films were grown under stoichiometric conditions with different doping densities which resulted in different carrier concentrations and disorder levels (as characterized by the residual resistivity ratios (RRR)) ~\cite{yue2022anomalous}.  bLaSTO73 (7.3 $\times$ 10$^{20}$/cm$^3$; RRR = 20) and bLaSTO18 (1.8 $\times$ 10$^{20}$/cm$^3$; RRR = 20) were grown with a SrTiO$_3$ buffer layer whereas the LaSTO72 (7.2 $\times$ 10$^{20}$/cm$^3$; RRR = 6), LaSTO27 (2.7 $\times$ 10$^{20}$/cm$^3$; RRR = 14), and NdSTO25 (2.5 $\times$ 10$^{20}$/cm$^3$; RRR = 12) were grown directly on the LSAT(001) substrate. This gave a range of carrier concentrations that span 1.79-7.29 $\times$ 10$^{20}$/cm$^3$.  Further growth details of the films are provided in ~\cite{yue2022anomalous,Critical2013thickness}.  These films have been extensively characterized with linear THz spectroscopy~\cite{kumar2024anomalous}.  We use a now standard implementation of the THz 2D coherent spectroscopic (2DCS) technique with two LiNbO$_3$ crystals that produce pulses with variable delay~\cite{Mahmood2021,barbalas2025energy}.  Two pulses A and B with electric field strengths of 45 kV/cm and a center frequency 0.7 THz are incident on the sample, and the signal emitted from the sample is recorded as a function of time (t), and the separation between the pulses $A$ and $B$ as $\tau$. The THz nonlinear response of the sample is $E_{NL}(\tau, t) = E_{AB}(\tau, t)-E_A(\tau, t)-E_B(t)$, where $E_{AB}$ is the transmitted signal when both $A$ and $B$ pulses are present, and $E_A$($E_B$) is the transmitted signal with pulse $A(B)$ alone.

Fig.~\ref{Chi3} shows the THz nonlinear response of the LaSTO27 and LaSTO72 SrTiO$_3$ thin films.  The time traces of the THz nonlinear $E_{NL}$ $\tau$ = 0 response for LaSTO27 and LaSTO72 are shown in Fig.~\ref{Chi3}(a) and ~\ref{Chi3}(b), respectively. $E_{NL}$  is decreasing with increasing temperature for both films.  The Fourier transforms of the THz nonlinear response $E_{NL}(\omega)$ for LaSTO27 and LaSTO72 are presented in Figs.~\ref{Chi3}(c) and ~\ref{Chi3}(d), respectively. $E_{NL}(\omega)$ indicates that the spectral maximum is found at the incident pulse maximum (shown in the SM, Fig. S1). We show the THz nonlinear response for the other carrier densities in the range of $1.8-7.3\times 10^{20}/$cm$^3$ in SM (Figs. S2 and S3), which shows that the THz nonlinear response generally decreases with increasing carrier density with a similar temperature-dependent response for all the samples.  The integrated intensity (integrated from $0.2 - 2$ THz) of the THz nonlinear signal (SW$_{NL}$) decreases with increasing temperature as displayed in Fig.~\ref{Chi3spec}(a).

What is the origin of this large nonlinearity?  Since SrTiO$_3$ is close to a ferroelectric transition, one might expect a large nonlinear response from the soft phonon as observed in bulk materials~\cite{kozina2019terahertz}. However the soft phonon mode in doped-SrTiO$_3$ thin films on LSAT substrates is not so soft.  In thin films it has been observed around 2-3 THz~\cite{sirenko2000soft,marsik2016terahertz} and in these particular films only for the lowest doping of bLASTO18 was any hint of the phonon found around 2 THz~\cite{kumar2024anomalous}.  The nonlinear signal dependence on the carrier concentration (Fig.~\ref{Chi3spec}(b)) indicates that it likely arises from the mobile carriers. Moreover, the largest THz nonlinear response was observed in the low temperature regime where the dc electrical resistivity exhibits a $T^2$-dependence.  As non-parabolicity in a metallic band can give strong THz nonlinearity, and the band structure of SrTiO$_3$ consists of three nonparabolic metallic bands it is likely the nonlinear signal comes from the low energy electronic band structure of doped-SrTiO$_3$.

\begin{figure*}[t]
    \centering
    \includegraphics[width = 16cm]{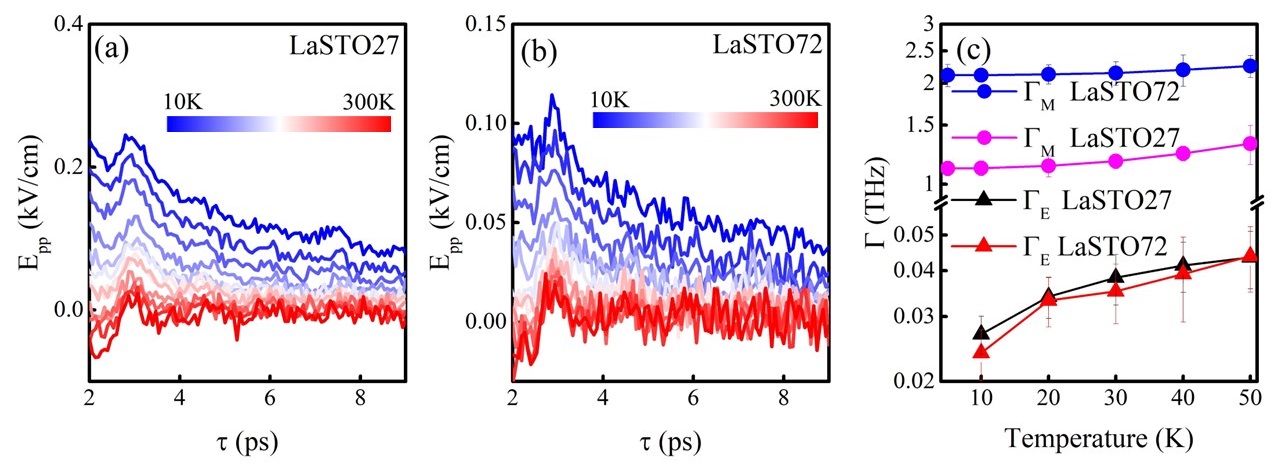}
    \caption{The nonlinear THz pump-THz probe response $E_{pp}(t\approx 0 \; \mathrm{ps}, \tau)$ of doped-SrTiO$_3$ thin films in the temperature range 5-300K at various concentrations. (a) LaSTO27 and (b) LaSTO72. (c) Energy relaxation rates ($\Gamma_{E}$) obtained from THz pump-THz probe response and momentum relaxation rates ($\Gamma_{M}$) extracted from optical Drude conductivity (measured using THz time-domain spectroscopy~\cite{kumar2024anomalous}) for the thin films. }
    \label{EPP}
\end{figure*}

To understand the $\chi^{(3)}$ process that contributes to the THz nonlinear response, we measured the full THz 2D coherent spectroscopic response as shown in Fig.~\ref{TwoD}(a,b) for the time traces for $t$ and $\tau>0.7$ps.  Data was taken with co-polarized $A$ and $B$ pulses.  From the positioning of the spectral intensity along the $\nu_\tau=0$ line in the Fourier transforms Figs.~\ref{TwoD}(c,d), we can see that the dominant contribution to the THz nonlinear response is from pump-probe $\chi^{(3)}$ process~\cite{barbalas2025energy} e.g. population decay.   Although one could analyze the frequency domain data directly, we have found it efficacious to analyze the time-domain data directly, which gives decay times averaged over the spectral bandwidth of the probe from a much faster 1D scan.  In Figs.~\ref{EPP}(a) and ~\ref{EPP}(b), we show the THz pump-THz probe response as a function of $\tau$ in the temperature range 10-120K for the LaSTO27 and LaSTO72 samples.  Note that here we plot the negative of the large negative peak near $t= 0$ ps as the pump pulse delay time $\tau$ was scanned.  The amplitude of the pump-probe signal $E_{PP}(t)$ is large at 5K and decreases with increasing temperature.  This response is largely independent of polarization as co- and cross-polarized responses show similar $\tau =0 $ responses (SM Fig. 9).

As discussed in our previous works~\cite{barbalas2025energy,bhandia2024anomalous} at long times, the pump-probe relaxation rate of metals is generally dominated by the energy relaxation rate ($\Gamma_E$), which is the rate that energy leaves the electronic system.  It can be contrasted with the (generally much faster) momentum relaxation rate ($\Gamma_M$) that is measured in the normal linear response optical conductivity, which is the rate the momentum leaves the electronic system.  To get insight into the pump-probe response, we fit the data at times longer than 2 ps using $ E_{NL}(t) =   A e^{-2\pi \Gamma_E t}$ where the $\Gamma_E$ is the energy relaxation rate. The time evolution of the pump-probe response is well characterized by a single exponential decay.  The obtained energy relaxation rates ($\Gamma_E$) are plotted in Fig.~\ref{EPP}(c).  $\Gamma_E$ is an increasing function of temperature up to 50K and is found to be roughly the same for both LaSTO27 and LaSTO72. 

For comparison, we have also plotted in Fig.~\ref{EPP}(c), the momentum relaxation rates ($\Gamma_{M}$) obtained from fits to the optical Drude conductivity measured using conventional THz time-domain spectroscopy measurements~\cite{kumar2024anomalous}. Note that we have included a factor of 2$\pi$ in the definition for the energy decay rate above so that the rates measured in time and frequency domains can be compared directly.  We find that $\Gamma_{M}$ is nearly 30-50 times larger than $\Gamma_E$ over the measured temperature  range. This indicates that the scatterings that remove energy from the electron system are largely not the same as those that lead to current/momentum decay.

The temperature dependence is also interesting.  In the conventional theory, the dominant channel for energy relaxation at low temperature is via coupling to acoustic phonons.  It is the expectation that $\Gamma_M > \Gamma_E$ because momentum loss can arise from non-momentum conserving electron-electron (umklapp and Barber) scattering and disorder scattering additionally.  It was predicted that at low temperature the energy relaxation rates goes as $\Gamma_E  = 12  \zeta(3)  \lambda \Big [ \frac{k_B T_e}{\hbar \omega_D} \Big ]^3  \omega_D$, where $\zeta$ is the Riemann zeta function and $\omega_D$ is the Debye frequency~\cite{allen_theory_1987}.  At higher temperatures  electron-phonon scattering becomes largely elastic and the energy relaxation rate is predicted to decrease as 1/T, giving a peak at a temperature of order the Bloch-Gr{\"u}neisen temperature.  Note this is the same temperature scale where the phonon contribution to resistivity is expected to crossover to linear in temperature in a conventional metal.   The same general considerations apply for optical phonons at temperatures above their characteristic energy scale~\cite{glorioso2022joule}. 

Our observations and these considerations give insight into the nature of the temperature dependence of the resistivity in doped SrTiO$_3$.  As discussed in Ref.~\cite{kumar2021quasiparticle} and above, it is has been proposed that the T$^2$ resistivity found at low temperature in doped SrTiO$_3$, arises not from electron-electron scattering, but from elastic scattering off {\it two} soft TO phonons at temperatures above their characteristic frequency.  In this equipartition regime, the resistivity will follow the square of the phonon density.  One expects only weak energy relaxation as the electron system gains energy through phonon scattering as often as it looses it.   Energy relaxation becomes increasingly inefficient as T increases and as discussed above $\Gamma_E$ is expected to decrease as 1/T (reflecting the $\gamma$T fermionic heat capacity) at temperatures above the characteristic maximum energy temperature for a phonon mode.  That we find $\Gamma_E$  increasing in the same temperature range that the resistivity goes as T$^2$, we believe rules out this scenario.   As discussed above, the coefficient of the T$^2$ resistivity smoothly evolves from a high density regime where multiple anisotropic bands are occupied to a low density regime where a single isotropic band is occupied.  We believe that it originates in electron-electron scattering, although it is still an open question how this is possible in a low density system with a small single isotropic Fermi surface.  These results align with our recent observation of a Fermi liquid like T$^2$ and $\omega^2$ dependent scattering rate with an anomalous scaling parameter~\cite{kumar2024anomalous} e.g. a low temperature state dominated by electron-electron interactions, albeit of an unconventional variety.

In conclusion, we have studied the THz nonlinear optical response of the dilute metallic phase in doped-SrTiO$_3$ thin films. We observed a large $\chi^{(3)}$ response in a temperature and doping regime where the dc electrical resistivity follows a T$^2$-dependence in doped-SrTiO$_3$.  We extracted the energy relaxation rate by fitting the THz pump-THz probe signal and compared it to the momentum relaxation rates.  The small scale of the energy relaxation rate compared to the momentum relaxation rate shows that scattering processes that degrade current largely do not carry away energy from the electronic system.  Moreover, the increasing energy relaxation as a function of T rules out coupling to a bath that is in the equipartion regime.  Among other aspects, this indicates that the T$^2$ resistivity in doped SrTiO$_3$ is likely to originate not in multi-phonon scattering but in electron-electron interactions.  It remains an open question how this is possible in a low density system with a small single isotropic Fermi surface.  We believe theoretical studies need to further consider how the electron-electron scattering is mediated through phonons or impurities.

We would like to thank K. Behnia, H. Hwang, D. van der Marel for helpful correspondences.  The project at JHU was supported by the NSF-DMR 2226666 and the Gordon and Betty Moore Foundation’s EPiQS Initiative through Grant No. GBMF9454. NPA had additional support by the Quantum Materials program at the Canadian Institute for Advanced Research. Synthesis and electrical transport (S.V. and B.J.) were supported by the U.S. Department of Energy through DE-SC0020211, and in part by the Center for Programmable Energy Catalysis, an Energy Frontier Research Center funded by the U.S. Department of Energy, Office of Science, Basic Energy Sciences at the University of Minnesota, under Award No. DE- SC0023464.  D.L. acknowledge support from the Air Force Office of Scientific Research (AFOSR) through Grant Nos. FA9550-21-1-0025. Parts of this work were carried out at the Characterization Facility, University of Minnesota, which receives partial support from the NSF through the MRSEC program under award DMR- 2011401. Film growth was performed using instrumentation funded by AFOSR DURIP award FA9550-18-1-0294.

\bibliography{main}


\clearpage

\widetext
\begin{center}
\textbf{\large Supplemental Material:  Absence of two-phonon quasi-elastic scattering in the normal state of doped--SrTiO$_3$ by THz pump-probe spectroscopy}
\end{center}
\setcounter{equation}{0}
\setcounter{figure}{0}
\setcounter{table}{0}
\setcounter{page}{1}

\section{Experimental Details}

Epitaxial La- and Nd-doped SrTiO$_3$ thin films were grown on (001) oriented (LaAlO$_3$)$_{0.3}$(Sr$_2$AlTaO$_6$)$_{0.7}$ (LSAT) single-crystal substrates using hybrid MBE approach. Elemental Sr and La were evaporated from effusion cells. Titanium was supplied as titanium tetra isopropoxide and oxygen from a radio-frequency plasma source operated at 250 W. Before growth, the substrates were in oxygen plasma in the MBE growth chamber at a substrate temperature of 900$^{\circ}$C for 20 min. Beam equivalent pressure ratios were set to be in the MBE growth window for stoichiometric films. The oxygen beam equivalent pressure was 5 $\times$ 10$^{-6}$ torr throughout the growth and cool down. The temperature of the La cell was varied to obtain the desired doping concentration at a constant growth rate of $\sim$ 160 nm/h. All films grew initially in a layer-by-layer growth mode that transitioned to step-flow mode. All substrates after film growth and undoped films were confirmed to be insulating. Here, we have grown the thin films of carrier concentration in the range of 1.79-7.29 $\times$ 10$^{20}$/cm$^3$. Further, bLaSTO73 (7.29 $\times$ 10$^{20}$/cm$^3$, 100 nm thickness) and bLaSTO18 (1.79 $\times$ 10$^{20}$/cm$^3$, 100 nm thickness) were grown with 20 nm SrTiO$_3$ buffer layer where as the LaSTO72 (7.17 $\times$ 10$^{20}$/cm$^3$, 66 nm thickness), LaSTO27(2.71 $\times$ 10$^{20}$/cm$^3$, 100 nm thickness), and NdSTO25 (2.5 $\times$ 10$^{20}$/cm$^3$, 194 nm thickness) were grown directly on the LSAT(001) substrate. Here, La doped SrTiO$_3$ thin films are indicated as LaSTO, La doped buffer layer films are denoted as bLaSTO, and Nd doped SrTiO$_3$ thin film denoted as NdSTO followed by the carrier concentration obtained from Hall transport measurements.   The films were different thicknesses as mentioned in text.  All thicknesses were below 180 nm to fully strain the films~\cite{Critical2013thickness}.

\section{Supplemental Results}

For THz nonlinear spectroscopic measurements we used two LiNbO$_3$ crystals in the tilted pulse front geometry as a source of THz generation in our system. The time traces of the THz signal from LiNbO$_3$ is shown in Figure S1 (a) and (b), respectively. The frequency spectrum of the signal in Figure S1 (a) and (b) are plotted in Figure S1 (c) and (d), respectively.  A and B THz pulses are nearly identical with a spectral maximum around 0.7 THz.

We implemented THz 2D coherent  spectroscopy to obtain the nonlinear $\chi^{(3)}$ response of the free carriers in doped-SrTiO$_3$ samples. Here we used two separate LiNbO$_3$ crystals generate the THz pulses of electric field strength $\sim$ 50 kV/cm. The THz pulses were detected using electro-optic sampling in GaP nonlinear crystal of thickness 0.5 mm. In general, 2D spectroscopic response is obtained as a function of gate time (t) and delay time ($\tau$). We used our setup in this work in at least three different modes to explore the nonlinear response.  The first is a 2 pulse experiment where the delay time is set at $\tau$ = 0 i.e. two pulses are overlapped. It includes contributions from all kinds of $\chi^{(3)}$ responses. In the second, the pump-probe response is obtained by fixing the gate time $t$ (typically to the max intensity which we take to be $t=0$)  and scanning the delay time $\tau$. Finally, the full 2D coherent spectra can be obtained by continuously varying the delay times. Most experiments are performed with co-polarized light.

The nonlinear $E_N(t,\tau$ = 0) response of doped-SrTiO$_3$ thin films at various carrier densities bLaSTO18, NdSTO25, LaSTO27, LaSTO72, and bLaSTO73 shown in Fig.~S2. The maximum nonlinear response ($E_{NL}(\tau=0,t)$) was observed for the least carrier density sample, i.e., bLaSTO18 as displayed in Fig.~S1(a). $E_{NL}(t,\tau=0)$ decreases with increasing  temperature as discussed in the main manuscript. As the carrier density increases, the nonlinear response also decreases as observed for NdSTO25, LaSTO27, LaSTO72, and bLaSTO73.

\begin{figure*}
    \centering
    \includegraphics[width = 12cm]{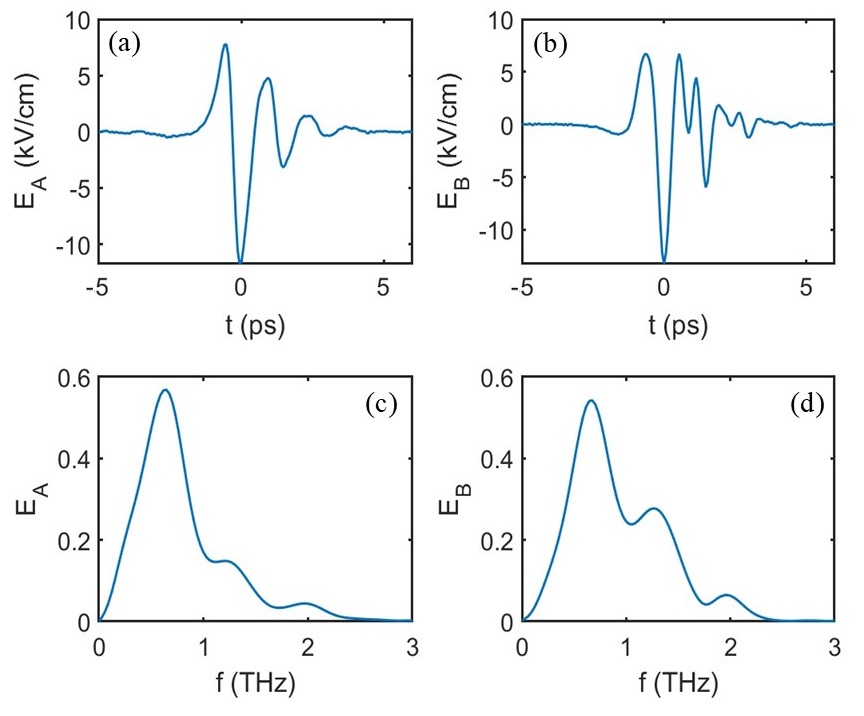}
    \caption{The THz frequency spectrum of the free space (a) A pulse and (b) B pulse. The corresponding frequency spectrum is plotted in (c) A pulse and (d) B pulse.}
    \label{Fig.1}
\end{figure*}

\begin{figure}
    \centering
    \includegraphics[width = 16cm]{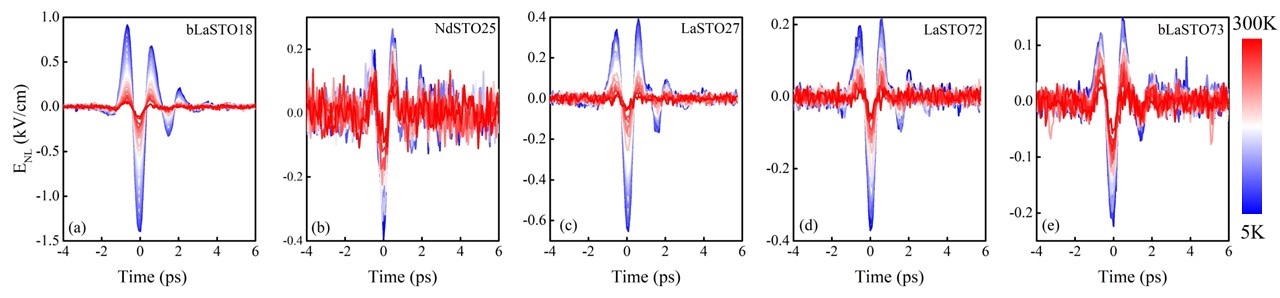}
    \caption{The time traces of the THz nonlinear signal $E_{NL}(t, \tau = 0)$ for various carrier densities.  (a) bLaSTO18, (b) NdSTO25, (c) LaSTO27, (d) LaSTO72, and (e) bLaSTO73.}
    \label{Fig.1}
\end{figure}

\begin{figure}
    \centering
    \includegraphics[width = 16cm]{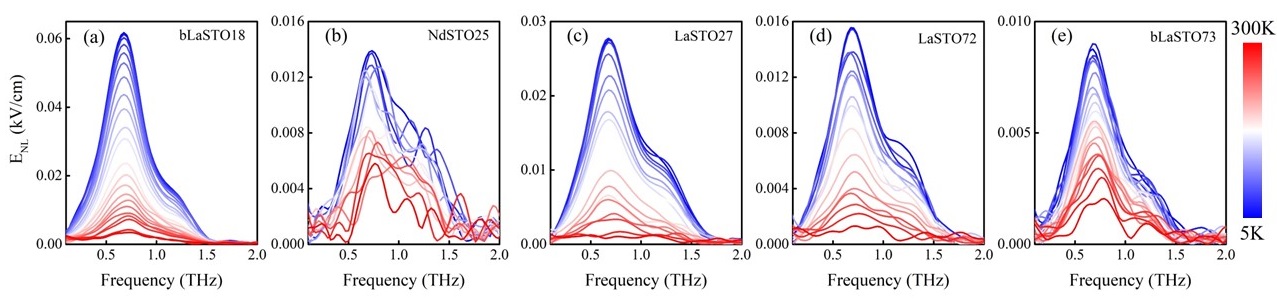}
    \caption{The Fourier transform of time traces shown in Figure 2.  (a) bLaSTO18, (b) NdSTO25, (c) LaSTO27, (d) LaSTO72, and (e) bLaSTO73.}
    \label{Fig.1}
\end{figure}

\begin{figure*}
    \centering
    \includegraphics[width = 10cm]{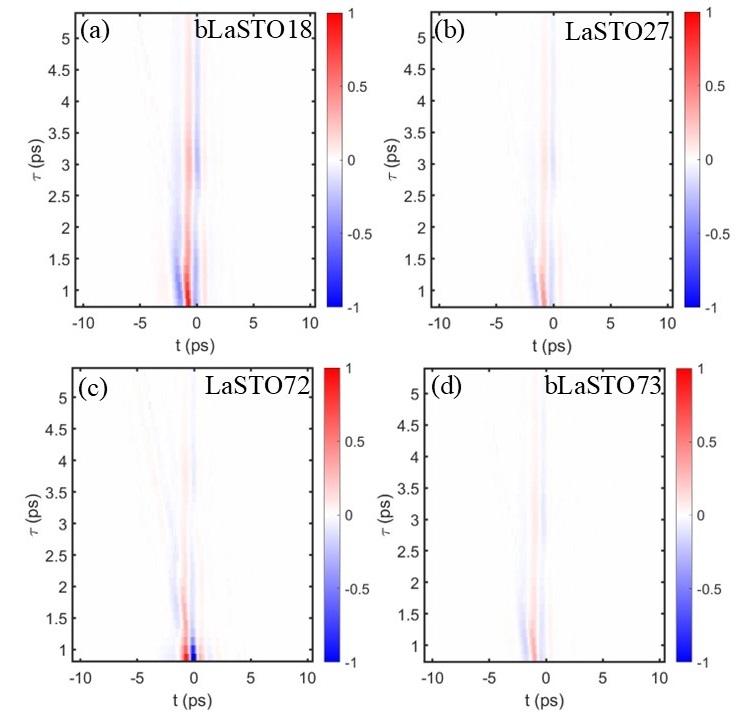}
    \caption{The time traces of the 2D THz nonlinear signal $E_{NL}(t, \tau )$ for various carrier densities in doped-SrTiO$_3$ thin films.  (a) bLaSTO18, (b) LaSTO27, (c) LaSTO72, and (d) bLaSTO73.}
    \label{Fig.1}
\end{figure*}

The Fourier transform of the Figure S2 data $E_{NL}(\omega)$ is shown in Figure S3. The frequency spectrum for all the sample is close to that in the incident pulses $E_{A}$ and $E_{B}$ (Fig. S1).  To get insight into the source of the nonlinear response, we implemented THz 2D coherent spectroscopy. The 2D THz spectra can be constructed by measuring the signal as a function of time ($t$) for variable time delay ($\tau$) between the $A$ and $B$ pulses.  The 2D nonlinear spectra for doped-SrTiO$_3$ thin films is shown Figure S4. In general, 2D spectra can be constructed for two configurations of $AB$ ($A$ pulse incident before $B$ pulse) and $BA$ ($B$ pulse incident before $A$). Here, we present the data for the $AB$ configuration.

\begin{figure}
    \centering
    \includegraphics[width = 10cm]{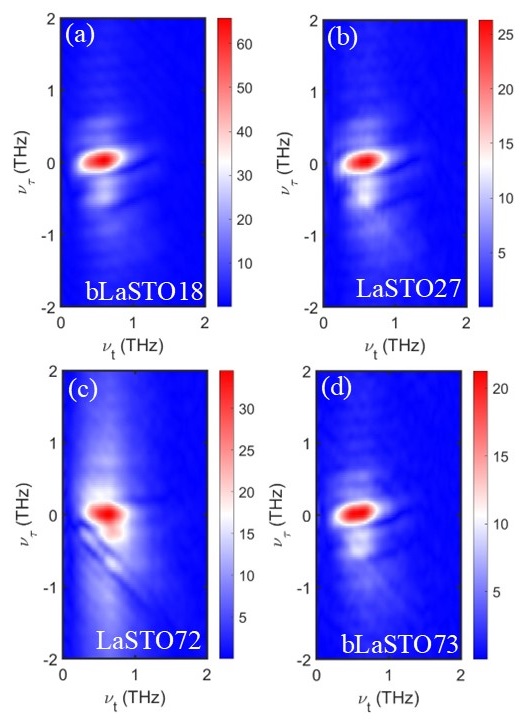}
    \caption{The Fourier transform of 2D THz spectra for Figure 4.  (a) bLaSTO18, (b) LaSTO27, (c) LaSTO72, and (d) bLaSTO73.}
    \label{Fig.1}
\end{figure}

 The 2D FFT of the nonlinear signal in Figure S4 is shown in Figure S5. Here, we considered only the $\tau>0.7$ ps values to exclude artifacts that are prominent as $A$ pulse and $B$ pulse overlap. Based on the position of peaks in the 2D spectra, we can -- in principle -- distinguish various nonlinear process such as pump-probe, rephasing and non-rephasing, harmonic signals as discussed elsewhere~\cite{kuehn2011two,Mahmood2021,barbalas2025energy}.  Here, we observed only the pump-probe ($\omega_t = \omega_A-\omega_B+\omega_B = \omega_B$) nonlinear signal as may be expected for a metal at low frequencies.

\begin{figure}
    \centering
    \includegraphics[width = 16cm]{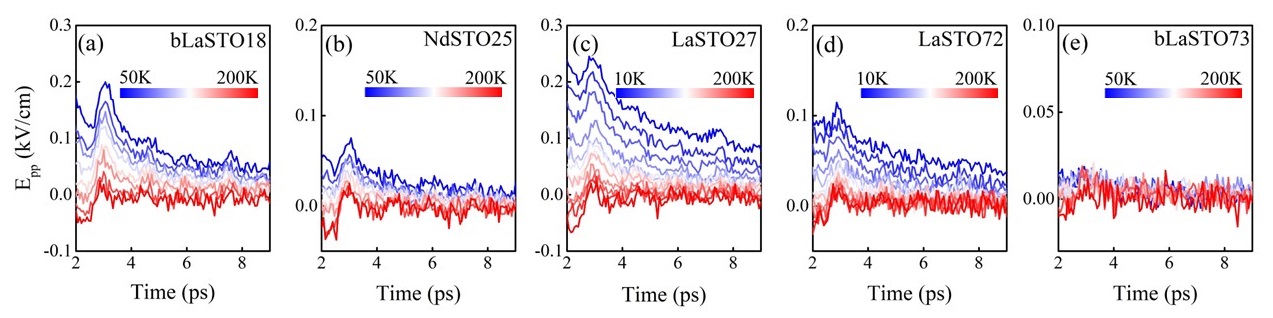}
    \caption{The nonlinear THz pump- THz probe time traces for various carrier densities in doped-SrTiO$_3$ thin films.  (a) bLaSTO18, (b) NdSTO25, (c) LaSTO27, (d) LaSTO72, and (e) bLaSTO73.}
    \label{Fig.1}
\end{figure}

As discussed in main text, we employed pump-probe measurements on doped-SrTiO$_3$ thin films.  These 1D scans are much faster than the full 2D scans and give relaxation rates that are an average over the pulse's bandwidth.  They give similar information if the relaxation rates are not strongly frequency dependent.  Here, we fixed the probe time (t) and varied the time delay between the pulses ($\tau$). Figure S6 shows the time traces of pump-probe signal for doped-SrTiO$_3$ thin films at various concentration. Here, we presented data from the different carriers concentrations.   Fits to this data give the energy relaxation rate as discussed in the manuscript.

\begin{figure}
    \centering
    \includegraphics[width = 16cm]{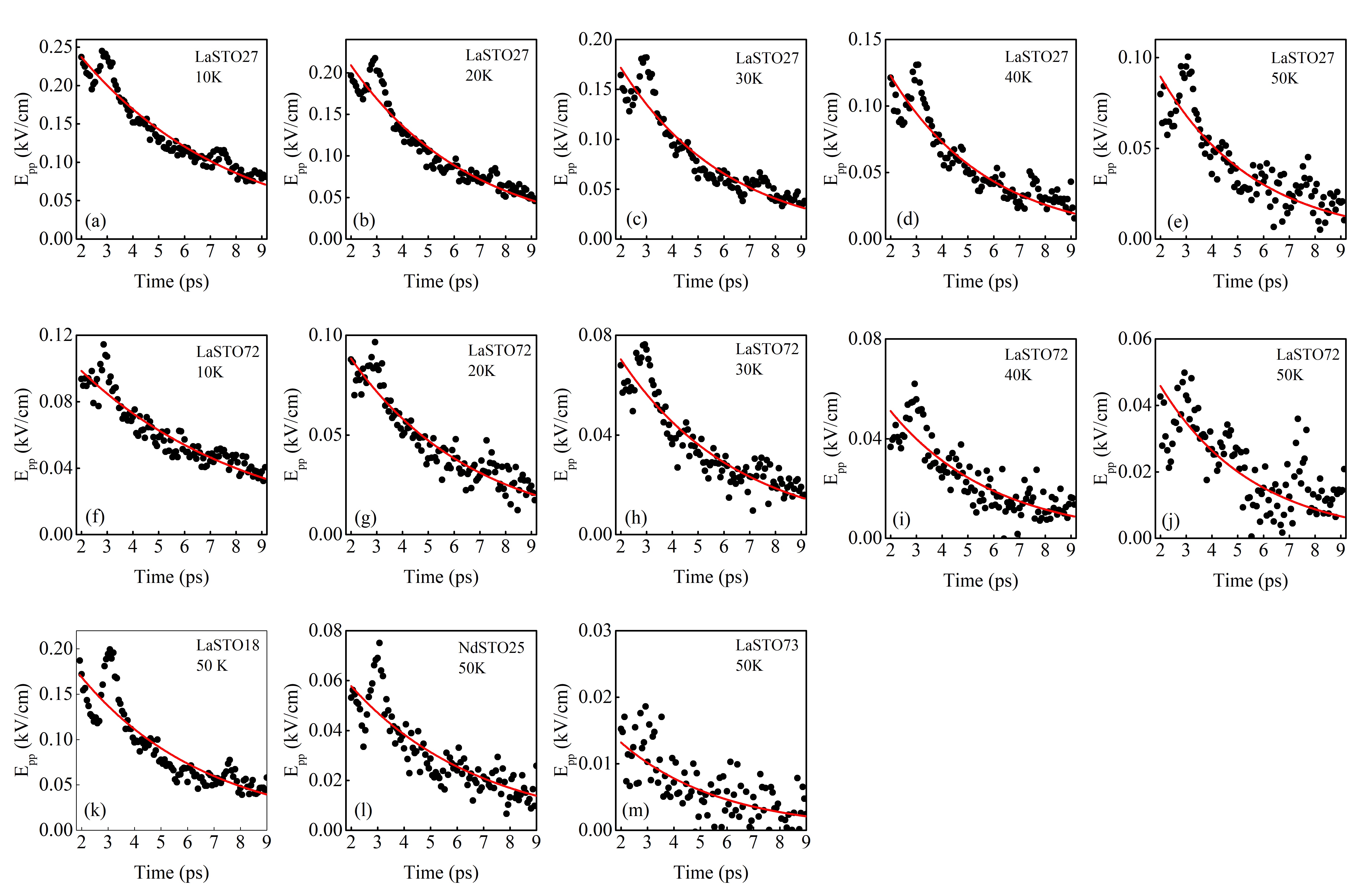}
    \caption{The nonlinear THz pump-THz probe time traces with an exponential fit for various carrier densities in doped-SrTiO$_3$ thin films. The solid circles are the experimental data, and the red line is an exponential fit at times longer than 2 ps.  ($A e^{-2\pi \Gamma_E t}$ where the $\Gamma_E$ is the energy relaxation rate).  Temperature-dependent THz pump-THz probe data for LaSTO27 (a) 10K, (b) 20K, (c) 30K, (d) 40K, (e) 50K. Temperature-dependent THz pump- THz probe data for LaSTO72 (f) 10K, (g) 20K, (h) 30K, (i) 40K, (j) 50K. (k) bLaSTO18 (l) NdSTO25, and (m) bLaSTO73 at 50K.}
    \label{Fig.1}
\end{figure}

\begin{figure}
    \centering
    \includegraphics[width = 6cm]{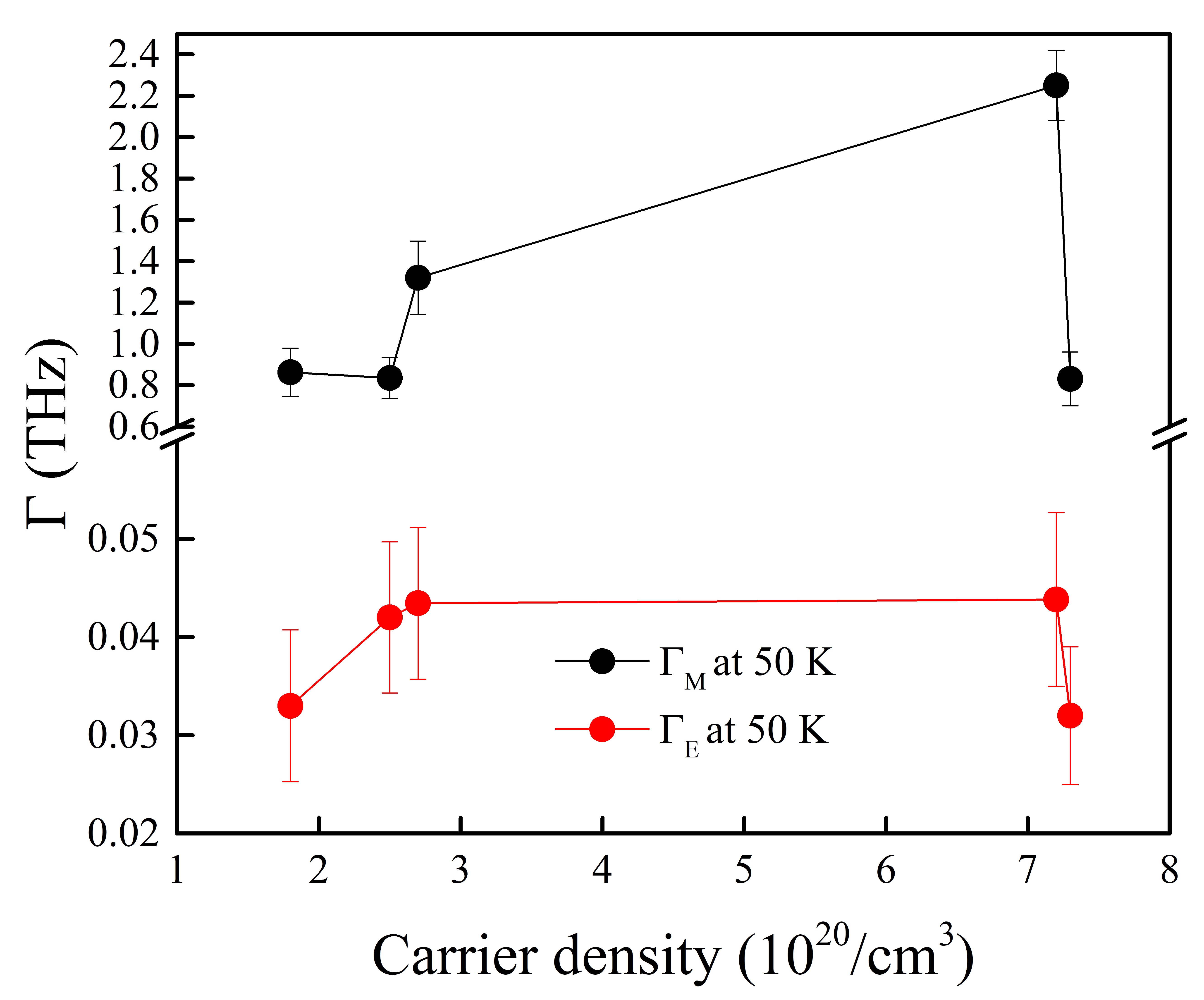}
    \caption{Carrier density dependent $\Gamma_M$ (momentum relaxation time) and $\Gamma_E$ (energy relaxation time) at 50K for doped-SrTiO$_3$ thin films. }
    \label{Fig.1}
\end{figure}

\begin{figure}
    \centering
    \includegraphics[width = 12cm]{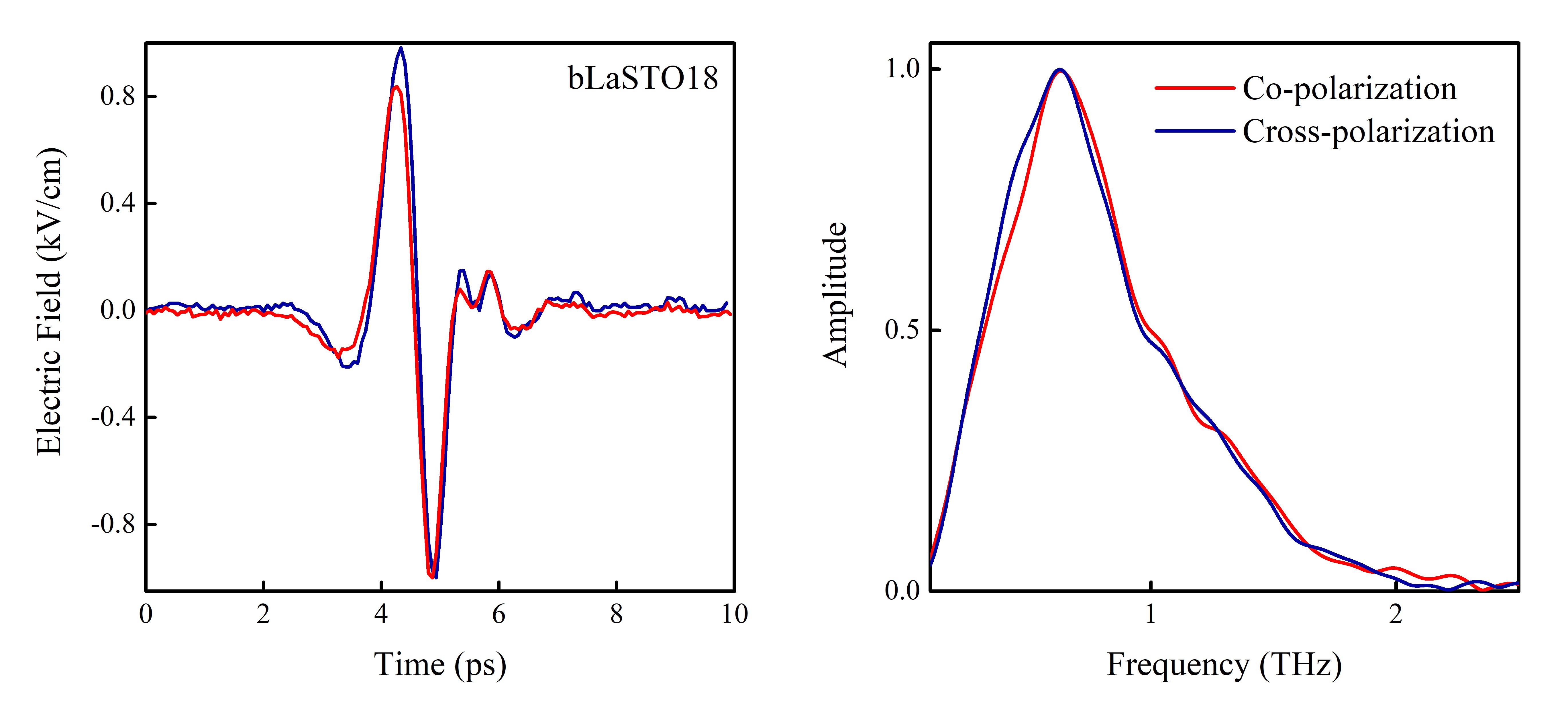}
    \caption{The co- and cross-polarization $\tau =0$ spectra of the nonlinear response for bLaSTO18 thin film. }
    \label{CrossCo}
\end{figure}

In Figure S\ref{CrossCo}, we show $\tau = 0$ spectra for experiments with co- and cross-polarized nonlinear responses for a bLaSTO18 thin film.  The almost uniform response gives strong indication that on the time-scales measured the response is dominated by uniform $A_{1g}$-like decays.   Uniform decays from larger to smaller momentum are only possible for simple Fermi surfaces by electrons loosing energy.  Hence this decay measures the energy relaxation rate.

\section{Free Carrier nonlinear Response}

Generally, intraband excitation of quasi-free carriers can be well represented by the Drude model.
\begin{equation}
    \varepsilon(\omega) = \varepsilon_{\infty}-\frac{\omega_{p}^{2} }{\omega^2 -i\omega \Gamma } 
\end{equation}

\begin{equation}
    \omega^2_{p} = \frac{ne^2}{m^*\varepsilon_{0}}, \; \; \; \; \; \;  \Gamma = \frac{e}{m^*\mu}   
\end{equation}
where $\varepsilon_{\infty}$ is the high-frequency permittivity, $\omega_{p}$ is the plasma frequency, $\Gamma$ is the scattering rate of the conduction carriers, $n$ is the carrier density, and $\mu$ is the mobility of carriers. The Drude model treats the effective mass of the conduction carriers constant which is the case for parabolic band dispersion.

One model to understand the free carrier nonlinear response of a material is as the modulation of the dielectric constant of free carriers by intense THz pulse.

\begin{equation}
   \epsilon (\omega) = \epsilon_D(\omega) \left[ 1-\frac{4 \pi n e^2}{m^*(E) \omega (\omega+i\Gamma (E))}  \right]
\end{equation}

where $m(E)$ and $\Gamma (E)$ are the energy dependent carrier effective mass and scattering rate. Since this free carrier dielectric constant contains energy dependent carrier mass and scattering rate, the observed nonlinear response can be due to energy dependent changes in the carrier effective mass which are present in non-parabolic band structures.

\end{document}